\PassOptionsToPackage{dvipsnames}{xcolor} 
\documentclass[proof]{WileyASNA-v1}

\articletype{Article Type}%

\received{31 July 2021}
\revised{17 October 2021}
\accepted{22 October 2021}

\newcommand\farcs{\hbox{$\,\!^{\prime\prime}$}}

\renewcommand\deg{^{\circ}}

\usepackage{natbib}
\begin{document}

\title{CO kinematics unveil outflows plausibly driven by a young jet in the Gigahertz Peaked Radio Core of NGC6328}

\author[1,2]{M. Papachristou}

\author[2,1]{K.M. Dasyra}

\author[1,3]{J.A. Fernández-Ontiveros}

\author[1,4]{A. Audibert}
\author[1,6]{I. Ruffa}

\author[7]{F. Combes}

\authormark{M. Papachristou \textsc{et al}}

\address[1]{\orgdiv{National Observatory of Athens (NOA)}, \orgname{Institute for Astronomy, Astrophysics, Space Applications and Remote Sensing (IAASARS)}, \orgaddress{\state{GR–15236}, \country{Greece}}}

\address[2]{\orgdiv{National and Kapodistrian University of Athens}, \orgname{Department of Astrophysics, Astronomy \& Mechanics, Faculty of Physics}, \orgaddress{\state{Panepistimioupolis Zografou, 15784}, \country{Greece}}}

\address[3]{\orgdiv{Istituto di Astrofisica e Planetologia Spaziali (INAF–IAPS)}, \orgaddress{\state{Fosso del Cavaliere 100, I–00133 Roma}, \country{Italy}}}

\address[4]{\orgdiv{Instituto de Astrof{\'i}sica de Canarias},  \orgaddress{\state{Calle Vía Láctea, s/n, E-38205 La Laguna}, \country{Spain}}}

\address[6]{\orgdiv{School of Physics \& Astronomy, Cardiff University}, \orgaddress{\state{Queens Buildings, The Parade, Cardiff CF24 3AA}, \country{UK}}}

\address[7]{\orgdiv{Observatoire de Paris, LERMA, Coll\`ege de France, CNRS, PSL Univ., Sorbonne Univ.}, \orgaddress{\state{F-75014 Paris}, \country{France}}}

\corres{\email{mpapaxristou@noa.gr}}

\abstract{We report the detection of outflowing molecular gas in the center of the nearby (z=0.014) massive radio galaxy NGC\,6328. The radio core of the galaxy, PKS\,B1718-649, is identified as a Gigahertz Peaked Spectrum source with a compact (2 pc) double radio lobe morphology. We used ALMA CO(2-1) and CO(3-2) observations at 100~pc resolution to study the gas kinematics up to $\sim 5\,\mathrm{kpc}$ from the galaxy center. 
While the bulk of the molecular gas is settled in a highly warped disk, in the inner $300\,\mathrm{pc}$ of the disk and along with the orientation of the radio jet, we identified high-excitation and high-velocity gas that cannot be attributed to any regular kinematic component based on our detailed 3D modeling of the ALMA data. 
The high-velocity dispersion in the gas also suggests that it is not part of an inflowing, shredding structure. These results suggest the presence of a molecular outflow of $3$ to $8$ solar masses per year. The outflow possibly originated from the interaction of the jet with the dense interstellar medium, even though the radio emission is detected closer to the center than the outflow. In this sense, this source resembles NGC\,1377, 4C31.04 and ESO\,420-G13, in which the outflows are linked to faint or past jet activity.}


\jnlcitation{\cname{%
\author{Papachristou M.}, 
\author{Dasyra K.M.}, 
\author{Fernández-Ontiveros J.A.}, 
\author{Audibert A.}
\author{Ruffa I.}, and 
\author{Combes F.}} (\cyear{2021}), 
\ctitle{CO kinematics unveil outflows plausibly driven by a young jet in the Gigahertz Peaked Radio Core of NGC6328}, \cjournal{Astron. Nachr}, \cvol{2017;00:1--6}.}

\maketitle

\section{Introduction}

A central question in galaxy evolution is the role and importance of the Active Galactic Nuclei (AGN) in the quenching of star formation caused as energy and momentum released by the AGN couples with the Interstellar Medium (ISM) and particularly the molecular gas phase, the fuel required to form new stars \citep{krumholzStarFormationLaw2009}.
Radio jets have been proposed as one of the most efficient mechanisms on providing this ISM-AGN coupling.
As shown by hydrodynamic simulations \citep{mukherjeeRelativisticJetFeedback2018}, young jets can drive expanding thermal bubbles (cocoons) as they propagate through the dense ISM creating shocks that disrupt molecular clouds and accelerate the gas through ram pressure. A class of radio sources associated with young compact jets are Gigahertz Peaked Spectrum (GPS) which characteristic peak in their radio spectrum can be caused by two possible mechanisms: the free-free absorption (FFA) of the synchrotron radiation by the ionized, shocked ISM or from the Synchrotron self-absorption (SSA) \citep{odeaCompactSteepspectrumPeakedspectrum2020}

The radio source PKS 1718-649 is a typical and well-studied GPS, peaking at 3.8 GHz. Its spectrum is best fitted with the FFA model, while high resolution ($\sim$0.3\,$\mathrm{pc}$) \textit{Very-Long-Baseline Interferometry} observations resolve the source into a bipolar structure, with the two lobes being separated by a projected distance of $\sim$2\,$\mathrm{pc}$ 
\citep{tingayNearestGHzPeakedSpectrum1997}

The host galaxy, NGC\,6328, is an early type galaxy at a distance of 62\,$\rm{Mpc}$ $(z=0.014313)$ 
making the radio source one of the closest GPS known. The stellar emission is dominated by a bright bulge with a typical S\'ersic index of $1/4$. A faint spiral structure is marginally seen extending up to 20~kpc in an almost face-on 
geometry. In the central 8~kpc, \textit{Hubble Space Telescope} imaging showed a dust lane tracing the outer spiral 
arms up to 4-5 kpc from the center, where it turns into an almost edge-on, north-south oriented disk. Atomic Hydrogen observations revealed a massive HI disk (1.1$\times10^{10}\,\mathrm{M}_\odot$) of maximum radius at $110\farcs$ (32\,$\mathrm{kpc}$), extending beyond the stellar continuum emission \citep{maccagni_what_2014}. 
The disk kinematics can be modeled with a warped disk of 220\,$\mathrm{kms^{-1}}$ circular velocity, which starts from <$30\farcs$ with inclination $i\approx 90\deg$ and position angle (PA) nearly parallel to the dust lane (180$\deg$) and warps to a face-on geometry ($\mathrm{PA}=110\deg,\, i=30\deg$) at a distance of $80\farcs$. 
\textit{Very Large Telescope} SINFONI K-band observations revealed warm $\mathrm{H}_2$ emission located in an inner disk or ring (within 700 pc) and two diametric hot spots 1\,kpc from the nucleus at a PA close to the orientation of the jet \citep{maccagniWarmMolecularHydrogen2016}. Also, X-ray observations from \textit{Chandra} and \textit{XMM-Newton} \citep{beuchertExtendedXrayEmission2018} showed evidence of an extended ($\sim$1\,kpc) ionized and hot gas component engulfing the warm molecular gas.

\section{ALMA data analysis and Modeling}
We present the analysis of archival ALMA observations of Band 6 CO(2-1) (PID 2015.1.01359.S) and Band 7 CO(3-2) (PID 2017.1.01638.S) in NGC\,6328.
The data were calibrated using the Common Astronomy Software Application (\textsc{CASA}) pipeline \citep{mcmullin_casa_nodate}. The continuum maps were made using natural weighting and have improved through phase and amplitude self-calibration. In both cases the radio source is unresolved. The CO line cubes were obtained after subtraction of the continuum in the line-free channels. The imaging was performed using natural weighting and channel widths of 20~km~s$^{-1}$, resulting in root-mean-square (rms) noise levels of 0.24 and 0.297\,mJy/beam for synthesized beams of 0\farcs27 $\times$0\farcs19 and 0\farcs35$\times$0\farcs29 for CO(2-1) and CO(3-2), respectively. 
We note that Band 7 observations were not optimized for the CO(3-2) transition and are only partially visible in the field-of-view (i.e.\,only CO(3-2) redshifted velocities were observed).

\begin{figure}[t]
	\centerline{\includegraphics[height=60mm]{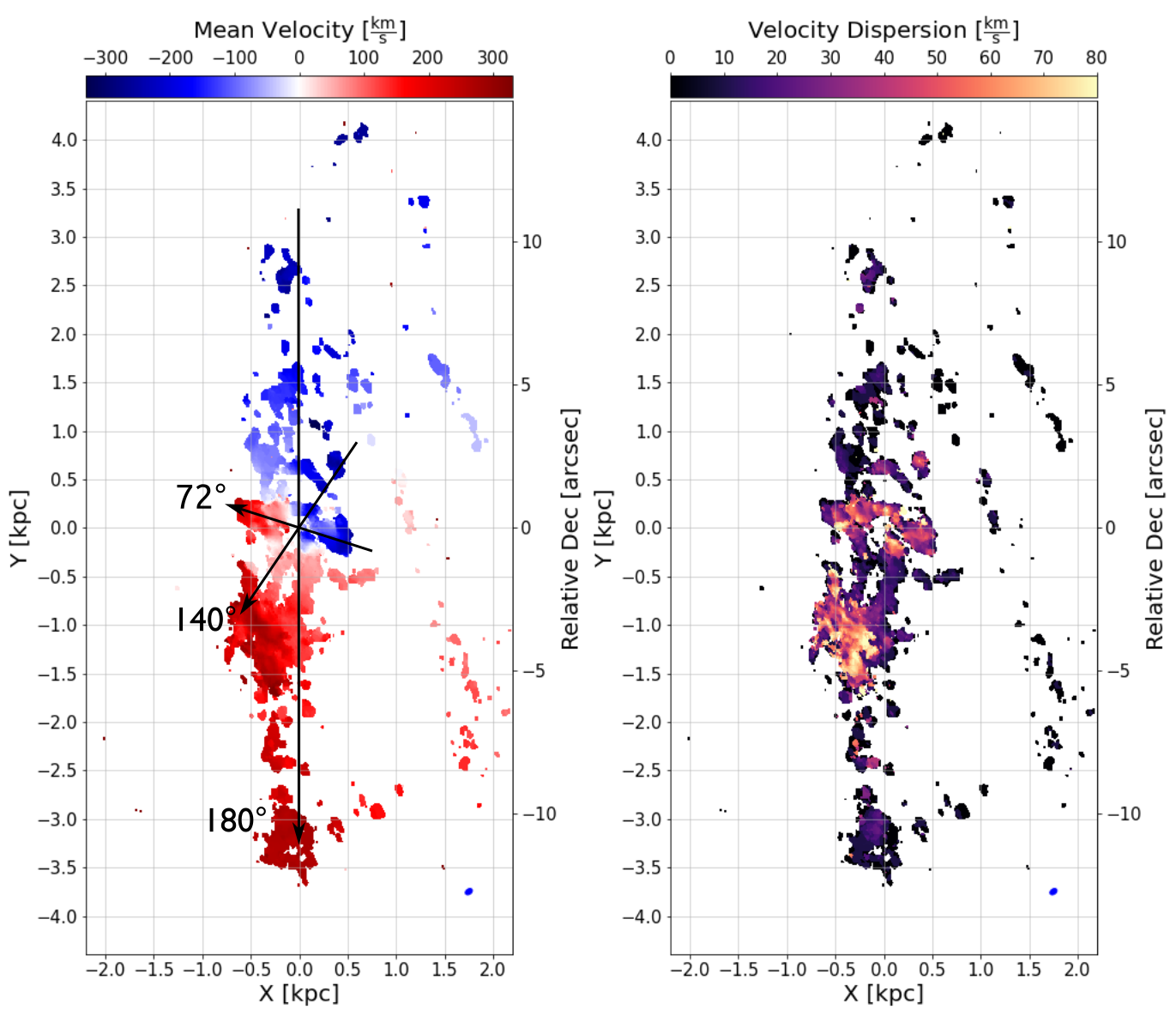}}
	\caption{First and second moments of the CO(2-1) cube. The radio continuum source is at (0,0). The position angles indicate the orientations of the inner disk (72$\deg$) and the larger kinematic component ($140-180\deg$).}
	\label{fig:moments}
\end{figure}

The molecular gas in NGC\,6328 is distributed in an asymmetric, highly inclined, and clumpy disk extending 
up to 4.5~kpc, closely correlated with the dust geometry. 
The $L'_\mathrm{CO(2-1)}$ luminosity calculated is 3.8$\times10^{8}\,\mathrm{K}\,\mathrm{kms^{-1}}\,\mathrm{pc}^2$ \citep{solomon_molecular_2005}, giving a total molecular gas mass of $1.8\times 10^{9} \,\mathrm{M}_\odot$ for a Milky Way conversion factor of 
4.6\,$\mathrm{M}_\odot\,(\mathrm{K}\,\mathrm{km s^{-1}}\,\mathrm{pc}^2)^{-1}$. 
CO(3-2) emission traces the same structure as CO(2-1) (at least in the observed receding side of the disk) and has a luminosity of $L'_\text{CO(3-2)}$=8.4$\times10^{7}\,\mathrm{K}\, \mathrm{km s^{-1}}\, \mathrm{pc}^2$.

At least two kinematic components can be identified from the mean
line-of-sight (LOS) velocity (moment 1) map of the CO(2-1) emission (Fig.~\ref{fig:moments}). The first one is a central disk extending up to 650\,pc from the center with $i\approx$65$\deg$ and PA$\approx$72$\deg$. This component is nearly aligned with the warm $\mathrm{H}_2$ emission, and its projected velocity grows linearly reaching a maximum of $\sim$300\,$\mathrm{kms^{-1}}$.
A second, larger-scale kinematic component shows a possible warped geometry as the zeroth and maximum velocities correspond to a changing PA from 140$\deg$ at 1\,$\mathrm{kpc}$ to 180$\deg$ at the maximum extent of 4.5\,$\mathrm{kpc}$. The disk reaches maximum LOS velocities of $\sim$350\,$\mathrm{kms^{-1}}$ at a radius of 1-2\,$\mathrm{kpc}$. 
In the dispersion map we identify three areas of considerably higher velocity dispersion ($\sigma>$80\,$\mathrm{kms^{-1}}$) than the rest of the disk ($\sigma \approx$25\,$\mathrm{kms^{-1}}$). First an area in the central disk and second, two diametric regions at a PA=140-160$\deg$ at radii between 0.5 and 1.5\,kpc. The southern region is much more massive 
and extended than the northern one. These areas are cospatial with the $\mathrm{H}_2$ excited regions and consistent with the jet orientation. The CO(3-2) emission mostly shows distribution and kinematics similar to the CO(2-1). Differently from the CO(2-1), however, an inner 150\,$\mathrm{pc}$ CO(3-2) clump is detected with  velocities of 200\,$\mathrm{kms^{-1}}$ and $\sigma \simeq 45\,\mathrm{kms^{-1}}$, and it is shifted of 120\,pc south-east of the core at a PA of 140$\deg$. 

We modeled the regular kinematics of NGC\,6328 to understand the nature of its high velocity dispersion areas: could the observed velocity dispersion be due to projection effects of regularly rotating, non-coplanar disks or to dynamically unsettled gas due to inflows or outflows?
To tackle this question and better understand the (kinematic) gas properties, we developed a new 3D model (in the sky+velocity space) code that uses the tilted rings approximation and it has the advantage of providing a model physically consistent with the galaxy potential and other observational constraints. For the gravity potential we used 3 components: a Keplerian potential for the SMBH, a Hernquist potential for the bulge \citep{hernquist_analytical_1990} and the NFW potential for the dark matter halo \citep{navarroStructureColdDark1996}. We do not add the potential of the disk, as its contribution to the rotation curve is comparable to the error bar ($20 \,\mathrm{kms^{-1}}$).

\begin{figure}[t]
	\centerline{\includegraphics[width=80mm]{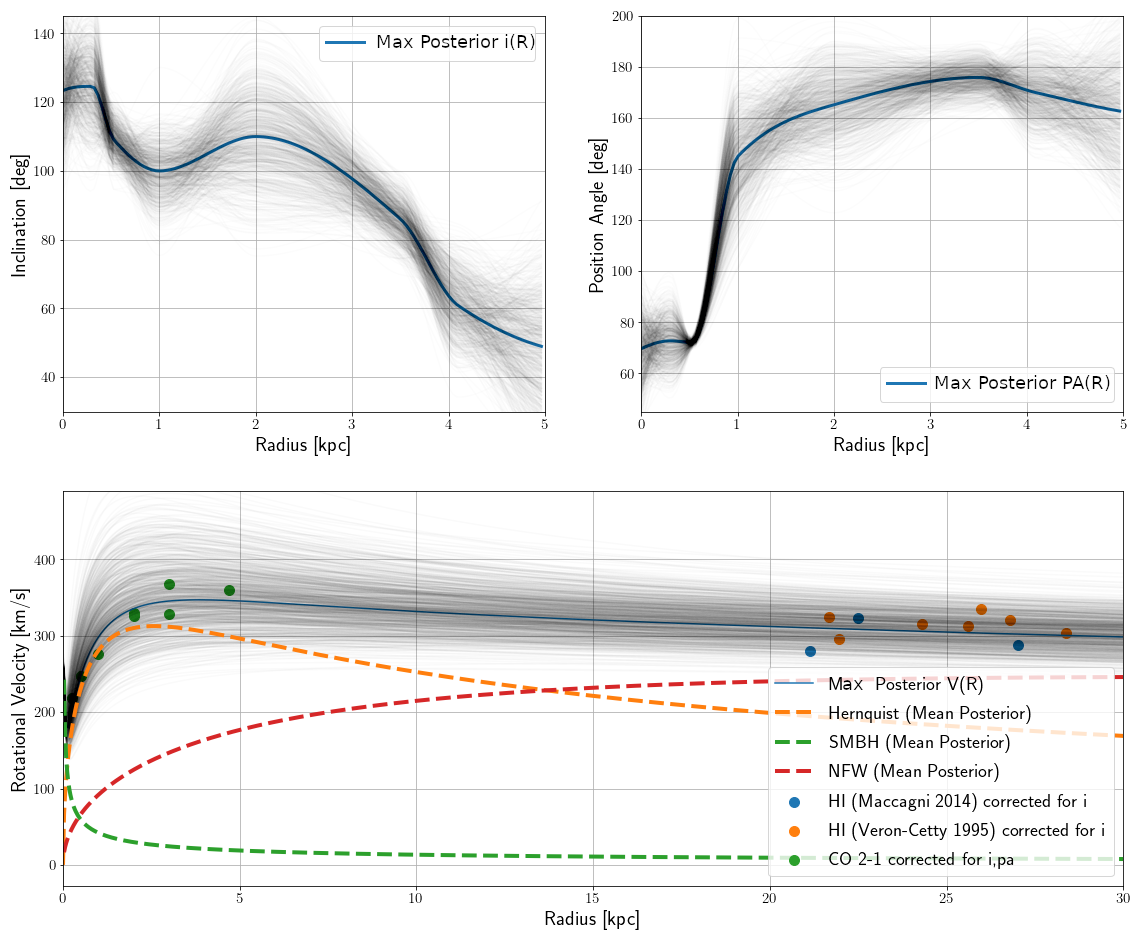}}
	\caption{Geometry (i, PA) (top) and Rotation Velocity (bottom) of the model sampled from the posterior distribution.}
	\label{fig:results}
\end{figure}

\begin{figure}[t]
	\centerline{\includegraphics[width=82mm]{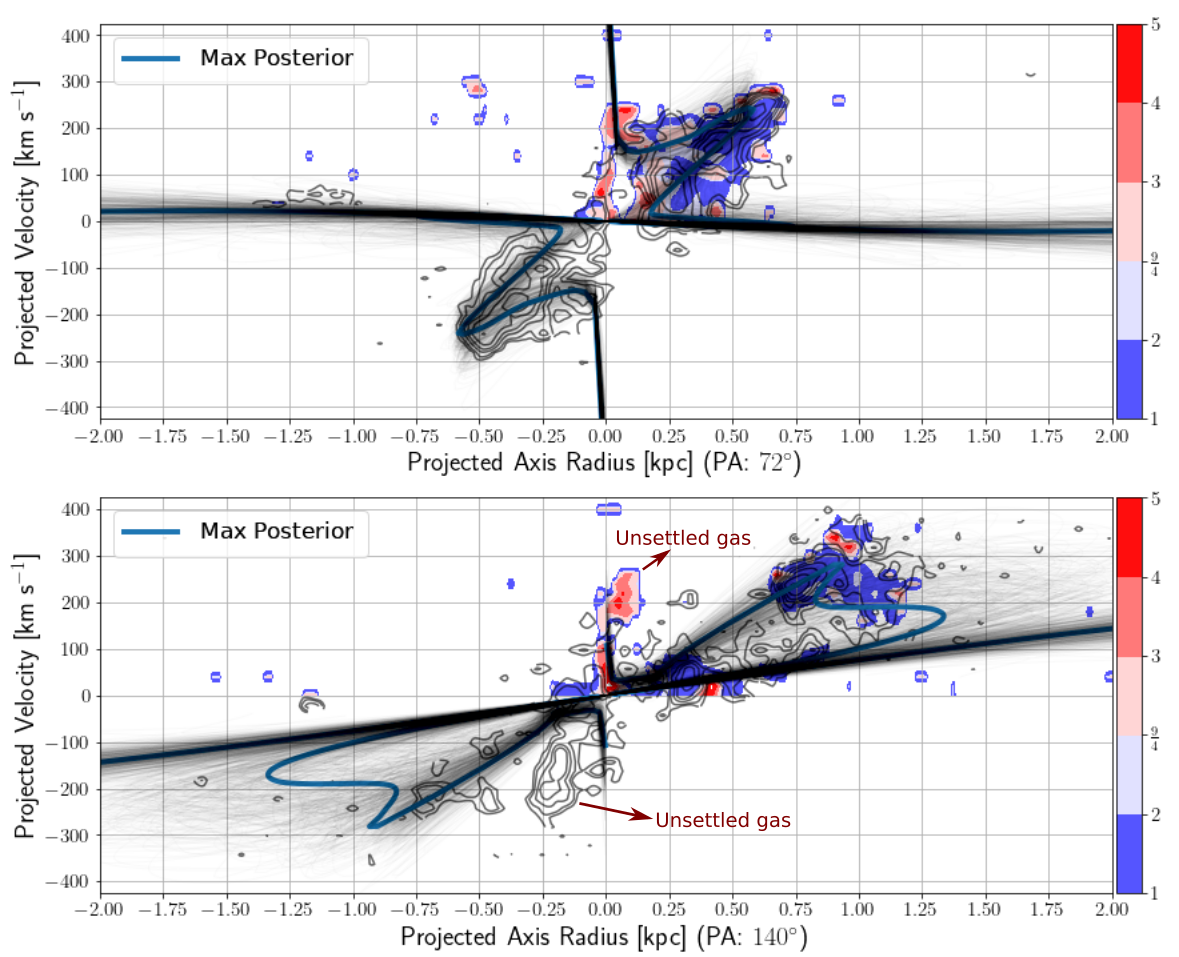}}
	\caption{Position Velocity plots of two characteristic PA, along the inner disk (72$\deg$) and along the jet orientation (140$\deg$). The colors represent $\mathrm{T}_\mathrm{CO(3-2)}/\mathrm{T}_\mathrm{CO(2-1)}$, where CO(3-2) has been detected, and the contours represent the CO(2-1) emission. The lines are the projected velocities of the tilted disks model. The maximum of the parameter posterior distribution are shown in blue and a sampled selection from the same distribution in black.}\label{fig:pvd}
\end{figure}

The model cube is created by iterating over a set of rings, which are differentiated by their PA, $i$, and rotational velocity. Next, we calculate the LOS projection of every ring in position and velocity and add an emission line in our model cube with an intensity matching that of the data. The gas velocity dispersion $\sigma$ is also parameterized into the model. $i$ and PA are modeled as smooth functions of radius through a monotonic cubic spline interpolation between a specific set of rings which are used as parameters. This parameterization choice facilitates the use of prior information for specific rings which we can infer from observations such as dust obscuration and the geometry of the HI disk. 
The parameter estimation is done by sampling, with the Markov Chain Monte Carlo sampler \textit{emcee} \citep{foreman-mackey_emcee_2013}, the posterior distribution of the parameters which is calculated by the product of the likelihood of our model and the joint prior distribution of the parameters. The likelihood is calculated as the combined probability of observing the data-cube given our model and the observed HI velocity data at the outskirts of the galaxy (\cite{maccagni_what_2014,veron-cetty_pks_1995}) given the model rotation curve.
The best-fitting set of parameters (i.e.\,the maximum of their posterior distributions (MAP), see Fig~\ref{fig:results}) can explain most of the high velocity dispersion regions of the molecular gas through projection effects (Fig.~\ref{fig:pvd}).

\begin{figure}[t]
	\centerline{\includegraphics[width=75mm]{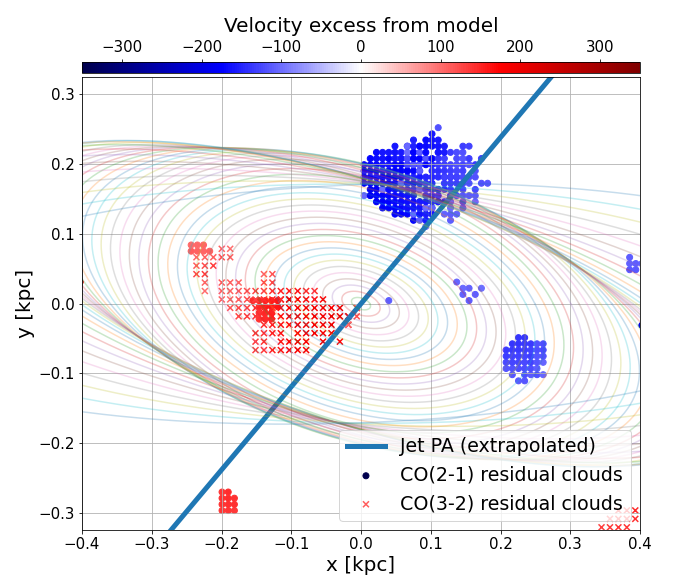}}
	\caption{Residual clouds of CO(2-1) \& CO(3-2) in the inner 500 pc with more than $100\,\mathrm{kms^{-1}}$ velocity excess from the regular rotation}\label{fig:residuals}
\end{figure}

However, velocity residuals are still evident for any acceptable smooth curve of PA and i. The most unambiguous residuals are those seen near the center of the galaxy (100-300\,$\mathrm{pc}$) at PA ranging from -30$\deg$ to -40$\deg$, and velocities ranging from 0 to $-250\,\mathrm{kms^{-1}}$ for the CO(2-1) and for the CO(3-2) clump we mentioned earlier at PA$\simeq 140\deg$ with velocity of $200\,\mathrm{kms^{-1}}$  (Fig~\ref{fig:pvd},~\ref{fig:residuals}).
The emission in these regions shows a steep rise in velocity that cannot be attributed to any possible projection effect of the warped disk. Any possible combinations of parameters capable of producing these projected velocities gives us highly unrealistic potential parameters and results that are inconsistent with the rest of the data. 

 \section{Discussion: Residuals and their link to an outflow}
As these clumps are 100-300\,pc from the nucleus, they are outside the sphere of influence of the SMBH ($<100\,\mathrm{pc}$).
Given the high velocity dispersion, implying a turbulent dynamical state of the unsettled gas, in the newly-examined CO(3-2) and CO(2-1) data (which show unsettled gas emission from -250 to 250\,$\mathrm{kms^{-1}}$), we argue that the non-rotating gas in emission cannot be attributed to the shredding of an individual infalling structure. Therefore, we suggest that we are observing an outflow. \cite{maccagniALMAObservationsAGN2018} observed CO(1-2) in absorption at $\sim$360$\mathrm{kms^{-1}}$ with a dispersion of $\sim$60$\mathrm{kms^{-1}}$ in front of the mm radio. In this paradigm, the gas in absorption can be an infalling cloud, part of the AGN feeding, or a fountain-like effect of previously accelerated outflow.
Another evidence supporting the outflow scenario is the excitation of the gas. We analyzed the gas excitation by comparing the brightness temperature ratio $\mathrm{T}_\mathrm{CO(3-2)}/\mathrm{T}_\mathrm{CO(2-1)}$, for the redshifted emission, to its LTE value of $\frac{9}{4}$ (Fig.~\ref{fig:pvd}) for optically thick gas. The regularly rotating, well-modeled gas is sub-thermally excited in contrast with the unsettled one. This increase can be due to extra density components or to the evaporation of outer cloud layers that makes them optically thin, or to the presence of extra excitation mechanisms (e.g., shocks, CRs) in the outflow. Similar results have been found for the excitation of jet-impacted clouds in the nearby galaxy IC\,5063 \citep{dasyra_alma_2016}. From the intensity of the high-velocity residuals, we conservatively estimate the mass of the outflow to be 3-4$\times10^6\,\mathrm{M}_\odot$ using an $a_\mathrm{CO}$ conversion factor of 0.8\,$\mathrm{M}_\odot(\mathrm{K}\, \mathrm{kms^{-1}}\,\mathrm{pc}^2)^{-1}$, and a mass outflow rate of 3-8\,$\mathrm{M}_\odot/\mathrm{yr}$. This leads to an outflow kinetic power between $2-7\times10^{40}\,\mathrm{ergs/s}$ and a momentum rate of $1-7\times 10^{33}\,\mathrm{dyn}$. A central star forming region would be unable to drive the outflow, as the SN kinetic power is less than $10^{40}\,\mathrm{ergs/s}$ and the stellar radiation pressure less than 5$\times10^{31}\,\mathrm{dyn}$.
Also, the low-luminosity AGN ($L_\mathrm{AGN}/c\approx 4.3\times10^{31}\,\mathrm{dyn}$) at the center of NGC\,6328 cannot provide the total radiation pressure needed to sustain the outflow momentum rate even if we adopt a high momentum boost factor of 20.

On the other hand, the jet power of $\sim 2\times 10^{43}\,\mathrm{ergs/s}$ \citep{wojtowiczJetProductionEfficiency2020} is 3 orders greater than the kinetic power of the outflow and can easily drive this outflow. The fact that the outflow is further away from the nucleus than the currently detected radio emission adds up to sources where the molecular wind traces a previously undetected radio jet as the cases of NGC1377 \citep{aaltoPrecessingMolecularJet2016}, 4C 31.04 \citep{zovaro_jets_2019} and ESO 420-G13 \citep{fernandez-ontiverosCOMolecularGas2020}.

\section*{Acknowledgments}
This project was funded by the  \fundingAgency{Hellenic Foundation for Research and Innovation} and the Greek \fundingAgency{General Secretariat for Research and Innovation} under grant number \fundingNumber{1882}
\begin{biography}{}{\textbf{M. Papachristou} is a PhD candidate at the National and Kapodistrian University of Athens under the supervision of prof. K.M. Dasyra. He is an employee at the National Observatory of Athens, working on the modeling of galactic kinematics for the detection of outflows.}
\end{biography}






%
\bibliography{proc_bibs}

\end{document}